\documentclass{article}

\usepackage{array}

\usepackage{arxiv}

\usepackage[utf8]{inputenc} % allow utf-8 input
\usepackage[T1]{fontenc}    % use 8-bit T1 fonts
\usepackage{hyperref}       % hyperlinks
\usepackage{url}            % simple URL typesetting
\usepackage{booktabs}       % professional-quality tables
\usepackage{amsfonts}       % blackboard math symbols
\usepackage{nicefrac}       % compact symbols for 1/2, etc.
\usepackage{microtype}      % microtypography
\usepackage{lipsum}		% Can be removed after putting your text content
\usepackage{graphicx}
\usepackage{natbib}
\usepackage{doi}

\title{Highly-sensitive measure of complexity captures Boolean networks regimes and temporal order more optimally}

\author{ \href{https://orcid.org/0009-0001-5335-9047}{\includegraphics[scale=0.06]{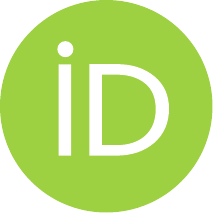}\hspace{1mm}Manuel de J. Luevano}\thanks{Use footnote for providing further
		information about author (webpage, alternative
		address)---\emph{not} for acknowledging funding agencies.} \\
	Unidad Académica de Física\\
	Universidad Autónoma de Zacatecas\\
	Zacatecas, Mexico \\
	\texttt{manuel.luevano@fisica.uaz.edu.mx} \\
	\And
	\href{https://orcid.org/0000-0003-1545-4315}{\includegraphics[scale=0.06]{orcid.pdf}\hspace{1mm}Alejandro Puga} \\
	Unidad Académica de Física\\
	Universidad Autónoma de Zacatecas\\
 Calzada Solidaridad esq. Paseo, La Bufa s/n C.P. 98060\\
 Zacatecas, Mexico \\
	\texttt{apuga@fisica.uaz.edu.mx
} \\
}

\renewcommand{\shorttitle}{\textit{arXiv} Template}

\hypersetup{
pdftitle={Highly-sensitive measure of complexity captures Boolean networks regimes and temporal order more optimally},
pdfsubject={q-bio.NC, q-bio.QM},
pdfauthor={David S.~Hippocampus, Elias D.~Striatum},
pdfkeywords={First keyword, Second keyword, More},
}

\begin{document}
\maketitle

\begin{abstract}
    In this work, several random Boolean networks (RBN) are generated and analyzed from two characteristics: their time evolution diagram and their transition diagram. For this purpose, its randomness is estimated using three measures, of which Algorithmic Complexity is capable of both a) revealing transitions towards the chaotic regime in a more marked way, and b) disclosing the algorithmic contribution of certain states to the transition diagram and their relationship with the order they occupy in the temporal evolution of the respective RBN. The results obtained from both types of analysis are useful for the introduction of both Algorithmic Complexity and Perturbation Analysis in the context of Boolean networks, and their potential applications in regulatory network models.
\end{abstract}

% keywords can be removed
\keywords{Random Boolean Networks \and Entropy \and Compressibility \and Algorithmic Complexity \and Perturbation Analysis}

\section{Introduction}
Discrete-time models are used in biology to capture the dynamic nature of biological systems. In particular, \textit{Boolean networks} are models that also assume discrete values for their variables, and allow us to have a qualitative view of the dynamics of a system. The origin of Boolean networks can be traced to Kauffman's work (\cite{KAUFFMAN1969437}), where he  introduces \textit{random genetic nets} to test the hypothesis that modern organisms are randomly assembled molecular automata. Later works, such as that of Kauffman (\cite{GLASS1973103}) where maps continuous biochemical control networks to their discrete analogue, and that of Thomas (\cite{THOMAS1973563}) who attempts to formalize genetic situations through models of complex control circuits, established the usefulness of Boolean network models to study biological regulatory systems.

Many studies highlight the usefulness of models of Boolean networks to analyze regulatory networks. Some examples are: a) the dynamic properties of the core network that controls the mammalian cell cycle (\cite{10.1093/bioinformatics/btl210}), b) simulation of the stomatal response in plants with a signal transduction network model (\cite{0262db1c2eb947029c310ea8c9c687f2}), c) virulence and response to pathogens resulting from the interactions between bacteria and the immune components of an infected host (\cite{ImmuneArticle}), d) the modeling of the \textit{lac} operon in \textit{Escherichia coli} (\cite{PMID:21563979}), and e) the modeling of the network of regulatory interactions that occur through various stages of embrionic development of the \textit{Drosophila} (\cite{ALBERT20031}). Most importantly, such type of models of  regulation and signalling serves to study a multitude of factors in an integrated way, rather than studying them in isolation.

In this paper, we present and compare various methodologies for analyzing the dynamics of some Boolean networks under the simplest of assumptions. First, we analyze the regimes of certain Boolean networks; then, we study the transitions between states in some simple Boolean networks applying perturbations. This work intends to introduce some useful tools for potential applications in the modeling and analysis of genetic regulatory networks, as dynamical systems with a causal regulatory network that can be found via gene perturbations.

\section{Background}
\begin{figure}[t]
    \centering
    \includegraphics[width=0.32\linewidth]{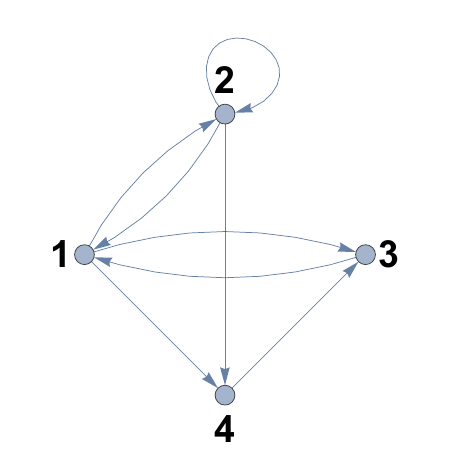}
    \includegraphics[width=0.32\linewidth]{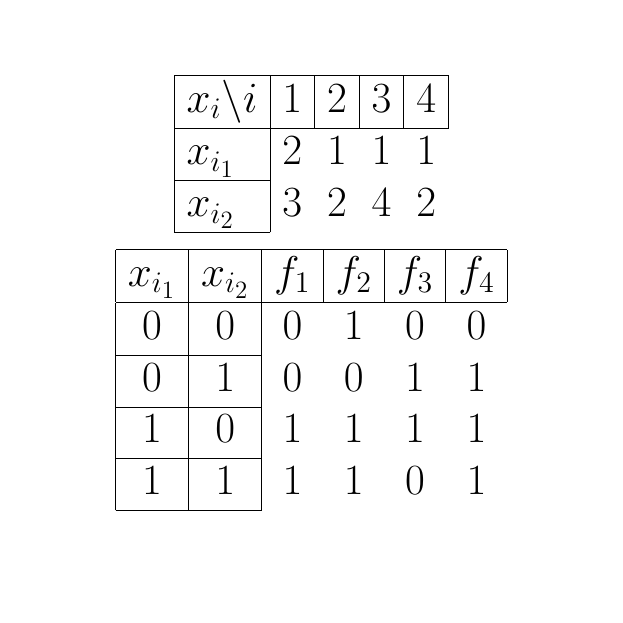}
    \includegraphics[width=0.32\linewidth]{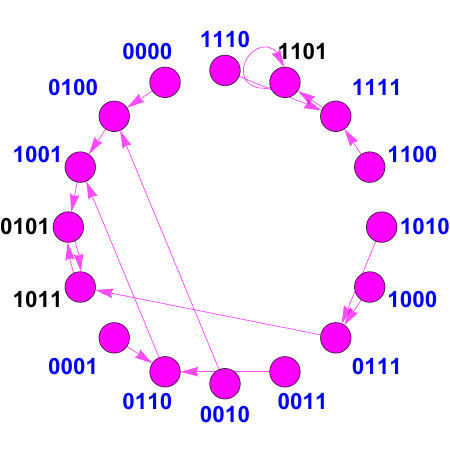}
    
    \caption{A specific Boolean network. Left: connectivity graph with $N=4$ and $k=2$. Medium: connectivity matrix and truth table. Right: transition diagram between all $2^{4}=16$ possible states, constructed from the connectivity matrix and the truth table; steady states or attractors are depicted in black.}
    \label{simple_RBN}
\end{figure}

Boolean networks, also known as $N$-$k$ networks, are characterized by two parameters: the number of nodes $N$ and the in-degree $k$. The state that each node can acquire is only one of two possible states -0 and 1, and is determined by a set of rules called \textit{boolean functions},

\begin{equation} \label{eq:2_1}
x_{i}\left( t+1 \right) = f_{i} \left( x_{j_1}\left( t \right),x_{j_2}\left( t \right),\cdots,x_{j_k}\left( t \right) \right)
\end{equation}

where the states $x_i$ of the nodes at time $t + 1$ are determined by the Boolean functions $f_i$ that depend on certain nodes at time $t$ (in this work, we make the assumption of \textit{synchronous update}; but there exist more flexible modeling frameworks that incorporate randomness in the duration of processes, see (\cite{CHAVES2005431})). The arguments of Boolean functions, as well as their order, are specified by a \textit{connectivity graph} and a \textit{connectivity matrix}. Boolean functions are represented by a \textit{truth table}. A Boolean network has a \textit{transition diagram} between all $2^{N}$ possible states (see Figure \ref{simple_RBN}).

\subsection{Random Boolean networks}

When Boolean functions are extracted from a random sample of 0s and 1s they become \textit{random boolean functions}, and the network becomes a \textit{random boolean network} (RBN) (\cite{RBN}). Boolean functions can be constructed with a uniform probability distribution. Or they can be constructed from a binomial probability distribution with parameter $p$, this parameter being therefore the probability that each of the Boolean functions returns a ``1'' as output. This parameter will also determine the type of system:

\begin{itemize}
    \item Chaotic, if $k\cdot 2p\cdot \left( 1-p \right) > 1$
    \item Ordered, if $k\cdot 2p\cdot \left( 1-p \right) < 1$
    \item At the edge of chaos, if $k\cdot 2p\cdot \left( 1-p \right) = 1$
\end{itemize}

\begin{figure}%[H]
    \centering
    \includegraphics[width=0.96\linewidth]{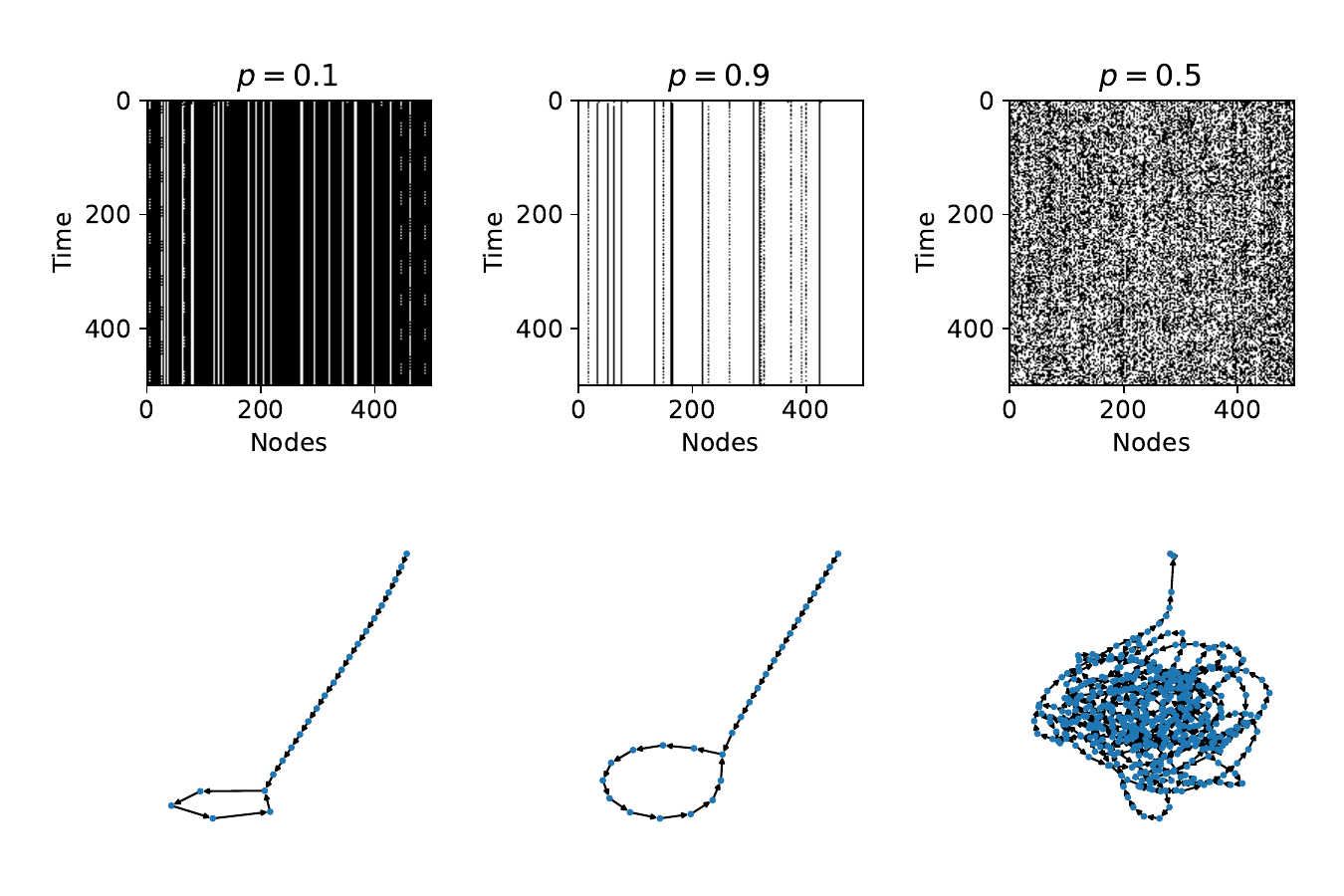}
    \caption{Time evolution diagrams (top) and their respective state transition graphs (bottom) for some RBNs with $N=500$ nodes and $k=5$. The RBNs share the same initial state and the same connectivity matrix, both generated randomly from uniform probability distributions.}
    \label{temporal_evolutions}
\end{figure}

In the last case, $k_c = \frac{1}{2p\cdot \left( 1-p \right)}$ is known as the \textit{critical in-degree}. Since these are probabilities, this is only true when the number of nodes $N$ is very large. Figure \ref{temporal_evolutions} shows time evolution diagrams for three RBNs with 500 nodes and in-degree $k=5$; two of them are in the ordered regime ($p=0.1$ and $p=0.9$) and the other one ($p = 0.5$) is in the chaotic regime. For the RBN with $p = 0.1$ there is a 90 percent probability that the Boolean functions deactivate their respective node; this can be confirmed in its time evolution diagram where there are many more deactivated nodes and in the rapid fall in a limit cycle attractor. Similarly, for the RBN with $p = 0.9$ there is a 90 percent probability that the Boolean functions activate their respective node, and a rapid fall in a limit cycle attractor is observed. On the other hand, it can be seen that the RBN with $p = 0.5$ is a chaotic system.

The time evolution and transition diagrams such as in Figure \ref{temporal_evolutions} allow us to analyze qualitatively the regimes in RBNs; for a more quantitative approach, we need a measure that is capable of capturing such regimes and the transitions between them. In the next section, we introduce some methods of measure on space-time patterns as candidates for the task at hand (for an alternative classification of chaotic/ordered RBNs, which is made from measurements on the basins of attraction of the state transition graphs, see (\cite{Wuensche_article})).

\section{Methodology}
\label{Meths}

The act of explaining what our eyes see has always been a human need. The act of trying to understand reality, which some have called science, common sense, etc., is nothing more than classifying randomness. When we discover the mechanism behind a phenomenon, we determine that it is not random and has a specific cause. To do this, a phenomenon can be abstracted to a chain of symbols and we call that ``the object''. However, measuring randomness is a challenge, and there are several measures at our disposal. In the next subsections, we review some of these measures.

\subsection{Information Theory and Entropy}

Information Theory (\cite{shannon}) is a branch of mathematics that deals with the quantification and communication of information. One of the fundamental concepts in information theory is the concept of entropy. In particular, the Shannon entropy is a measure of the amount of uncertainty or information content in a random variable.

The Shannon entropy $H\left( X \right)$ of a discrete random variable $X$ with probability mass function $p\left( x \right)$ is defined as

\begin{equation} \label{eq:3_1}
H\left(X\right) = -\sum_{x \in \mathcal{X}}p\left(x\right)\log_{2}p\left(x\right)
\end{equation}

where $\mathcal{X}$ is the set of all possible values of $X$. The logarithm is usually taken to base 2, so the entropy is measured in bits.

The Shannon entropy has a number of important properties, including:

\begin{itemize}
    \item  It is always non-negative, $H\left( X \right) \ge 0$.
    \item It is maximized when all the possible values of $X$ are equally likely, i.e., when $p\left( x \right) = 1/\left| \mathcal{X} \right|$ for all $x \in \mathcal{X}$. In this case, the entropy is $\log_2\left| \mathcal{X} \right|$.
    \item It is minimized when $X$ is a deterministic function of another random variable $Y$, i.e., when there exists a function $f$ such that $X = f\left( Y \right)$ with probability 1.
\end{itemize}

In the last case, $H\left( X|Y \right)=0$, since there is no uncertainty or information content in $X$ beyond what is already contained in $Y$.

Shannon entropy is a useful tool in a variety of applications, including coding theory, data compression, and cryptography (\cite{book:92629441}). The Shannon entropy indicates how much ``surprise'' or ``unexpectedness'' there is in the outcomes of a random variable. The higher the entropy, the more uncertain or unpredictable the random variable is, and the more information is needed to describe or communicate its outcomes. Conversely, the lower the entropy, the more certain or predictable the random variable is, and the less information is needed to describe or communicate its outcomes.

\subsection{Compression algorithms and Compressibility}

A \textit{lossless compression algorithm} is any encoding procedure that aims to represent a certain amount of information using or occupying a smaller space, making an exact reconstruction of the original data possible. An example is the Lempel-Ziv-Welch (LZW) algorithm (\cite{1055934}), a dictionary-based lossless compressor. The algorithm uses a dictionary to map the sequence of symbols to codes that take up less space. The LZW algorithm performs well, especially with files that have repetitive sequences. Another example of a lossless compression algorithm is the free and open-source file compression program called BZip2, which uses the Burrows-Wheeler algorithm (\cite{BZip2}), which only compresses individual files; It relies on standalone external utilities for tasks such as handling multiple files, encryption, and file splitting.

From the lossless compression of a piece of data we can define its \textit{compressibility rate} as the length of the codified data divided by the length of the original data. This is an intuitive measure of randomness because we expect that the more random a piece of data or string is, the more difficult to compress it.

\subsection{Algorithmic Complexity and BDM}

Program size complexity, also known as \textit{algorithmic complexity}, is a measure that quantifies algorithmic randomness. The algorithmic complexity (AC) or \textit{algorithmic information content} of a string $s$ is defined as the length of the shortest computer program $p$ that is executed in a prefix-free Universal Turing Machine and generates $s$ as an output (\cite{doi:10.1080/00207166808803030, 10.1145/321356.321363}):

\begin{equation} \label{eq:3_2}
K(s)=\min{\left\lbrace\left | p \right |:U(p)=s\right\rbrace}
\end{equation}

The difference between the original length and the minimum program size determines the complexity of data $s$. A string $s$ is said to be random if $K\left( s \right)$ (in bits) $\sim \left| s \right|$ (for a more intuitive measure of complexity that combines the notions of algorithmic information content and time, see the concept of Charles Bennett's logical depth (\cite{Bennett1992-BENLDA-10})). 

A technical drawback is that $K$ is not computable due to the Halting problem (\cite{copeland2004essential}), since we can't always find the shortest programs in a finite time without having to run all the computer programs, which means having to wait forever in case they never stop. Nevertheless, it is possible to randomly produce programs that produce a specific result to estimate its AC from above; for this purpose, the \textit{algorithmic probability} $m\left( s \right)$ of a binary string $s$ is defined as the sum over all (prefix-free)  programs with which a (prefix-free) Universal Turing Machine generates $s$ and stops:

\begin{equation} \label{eq:3_3}
m\left( s \right)=\sum_{p:U(p)=s}\frac{1}{2^{\left | p \right |}}
\end{equation}

The \textit{Coding Theorem Method} (CTM) (\cite{SOLOMONOFF19641, zbMATH03489017}) further establishes the connection between $m\left( s \right)$ and $K\left( s \right)$,

\begin{equation} \label{eq:3_4}
\left| -\log_2m\left( s \right)-K\left( s \right) \right| < c
\end{equation}

where $c$ is a fixed constant, independent from $s$. This theorem implies (\cite{calude2013information}) that the output frequency distribution of random computer programs to approximate $m\left( s \right)$ can be converted to estimates of $K\left( s \right)$ using

\begin{equation} \label{eq:3_5}
K\left( s \right)=-\log_{2}(m\left( s \right))+O\left( 1 \right)
\end{equation}

Finally, the \textit{Block Decomposition Method} (BDM) (\cite{e20080605}) is a powerful tool to estimate algorithmic complexity. This method allows the decomposition of a large system in smaller and more manageable pieces, facilitating the estimation of its total complexity. By dividing the data into fragments and precomputing CTM calculations on these fragments (\cite{delahaye2012numerical,SolerToscano2012CalculatingKC}), BDM provides a measure that lies between entropy and algorithmic complexity, allowing a better understanding of the structure and behavior of the system under study. In general, for $w$-dimensional objects, BDM is defined as

\begin{equation} \label{eq:3_6}
\textrm{BDM}\left( X,\left\{ x_{i} \right\}\right) = \sum_{\left( r_{i},n_{i} \right)\in \textrm{Adj}\left( X\right)_{\left\{ x_{i} \right\}} }\left[ \textrm{CTM}\left( r_{i} \right) + \log n_{i} \right]
\end{equation}

where the set $\textrm{Adj}\left( X\right)_{\left\{ x_{i} \right\}}$ is composed of the pairs $\left( r_{i},n_{i} \right)$, $r_{i}$ is an element from the decomposition of $X$ (specified by a partition $\left\{ x_{i} \right\}$, where $x_{i}$ is a submatrix of $X$) and $n_i$ is the multiplicity of each $r_{i}$.

%where $r_{i}$ is an element from the decomposition of $X$ (specified by the partition $\left\{ x_{i} \right\}$) and $n_i$ is the multiplicity of each $\left\{ r_{i} \right\}$.

\subsection{Perturbation analysis}

It is worth asking whether the randomness measures outlined above can serve us for more than simply determining how random an object is or how chaotic its temporal evolution is. Specifically, we can ask if a robust measure of randomness can help us characterize or even control a dynamic system.

The \textit{perturbation analysis} or \textit{reprogrammability analysis} is based on the change in the randomness of a dynamic system subjected to perturbations. Its purpose is to track in detail the changes in algorithmic information content (estimated by local observations) produced by natural or induced perturbations in evolving complex open systems. Algorithmic models can be derived from partial observations, and the algorithmic probability that these models produce the behavior of the system under analysis can be estimated. Perturbation analysis can also help us reconstruct space-time diagrams of dynamic systems (\cite{ZENIL20191160,Zenil_Kiani_Tegnér_2023}).

A simple example will allow us to illustrate the basic concepts of perturbation analysis. Let $G$ be a graph with a set of nodes $V\left( G \right)$ and a set of vertices $E\left( G \right)$. The dynamics of $G$ can be defined by transitions between different states. Let $G' = G \backslash e$ denote the operation of removing a vertex $e$ from $G$. The difference $\textrm{AID} := K\left( G \right) - K\left( G \backslash e \right)$ would be the shared mutual algorithmic information, or \textit{algorithmic information dynamics}, of $G$ and $G \backslash e$, where it is considered that $G$ is a time-dependent system and that $G'$ can be the result of the temporal evolution of $G$. We have the following cases:

\begin{itemize}
  \item $\left|\textrm{AID}\right|\le \log_{2}\left| V\left( G \right) \right|$. The vertex $e$ is contained in the algorithmic description of $G$, and can be recovered from the latter.
  \item $\left|\textrm{AID}\right|\sim \log_{2}\left| V\left( G \right) \right|$. The vertex $e$ does not contribute to the description of $G$. The relationship between $G$ and $G'$ is causal.
  \item $\left|\textrm{AID}\right| > \log_{2}\left| V\left( G \right) \right|$. $G$ and $G'$ do not share causal information. Removing $e$ results in a loss.
  \item $\left|\textrm{AID}\right| > \left| V\left( G \right) \right|$. There are two possibilities: a) $G'$ cannot be explained by $G$ alone, that is, $G'$ is not algorithmically contained in $G$, and therefore $e$ is a fundamental part or generative causal mechanism of $G$; and b) $e$ is not part of $G$ but it is noise.
\end{itemize}

In the first two cases, it is said that element $e$ is not essential in the explanation of G but it is an element that is produced by the natural course of evolution of $G$. In the last two cases, if $\textrm{AID} < 0$, element $e$ is said to make a high \textit{causal contribution} since its removal moves $G$ more toward randomness. And in the last case, if $\textrm{AID} > 0$, then the removal of element $e$ moves $G$ away from randomness (or towards simplicity), so it is unlikely that element $e$ is part of the generating mechanism of $G$, i.e., it is more likely that element $e$ is noise or part of another system.

In this work, a basic perturbation analysis will be carried out on the state transition graph of some RBNs using the randomness measures already outlined in the previous subsections.

\section{Results and Discussion}
\label{results}

In this section we present and discuss some results about the time evolution of RBNs whose connectivity graphs were generated like this: for each of the $N$ nodes, select one node $k$ times as an adjacent node following a uniform probability distribution over all $N$ nodes. We used the randomness measures of Section \ref{Meths} over boolean functions (truth tables) and time evolution diagrams. Entropy was calculated from the relative frequencies of 0s and 1s. For the calculation of compressibility, an array was first converted to a one-dimensional string concatenating the rows from top to bottom, and then such string was compressed using the LZW algorithm. Finally, Algorithmic Complexity for an array was estimated through the Block Decomposition Method (for the estimation of AC by BDM we used the \texttt{pybdm} library (\cite{pybdm}))

%This section may be divided by subheadings. It should provide a concise and precise description of the experimental results, their interpretation as well as the experimental conclusions that can be drawn.

\subsection{RBNs with $N=500$ nodes}
\begin{figure}
    \centering
    \includegraphics[width=0.64\linewidth]{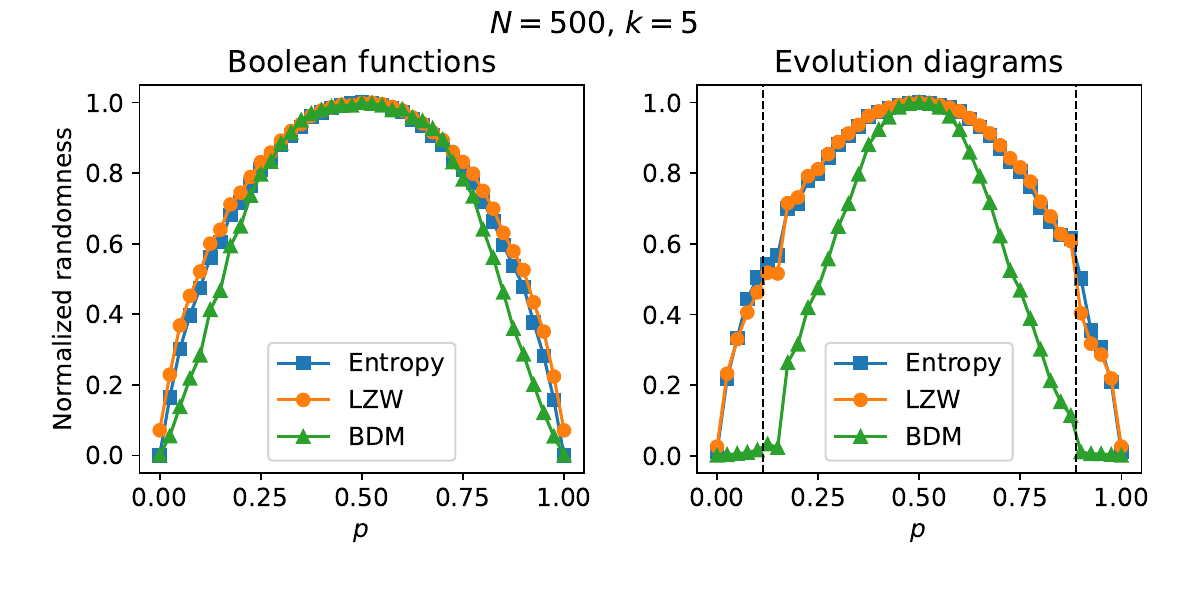}
    \includegraphics[width=0.64\linewidth]{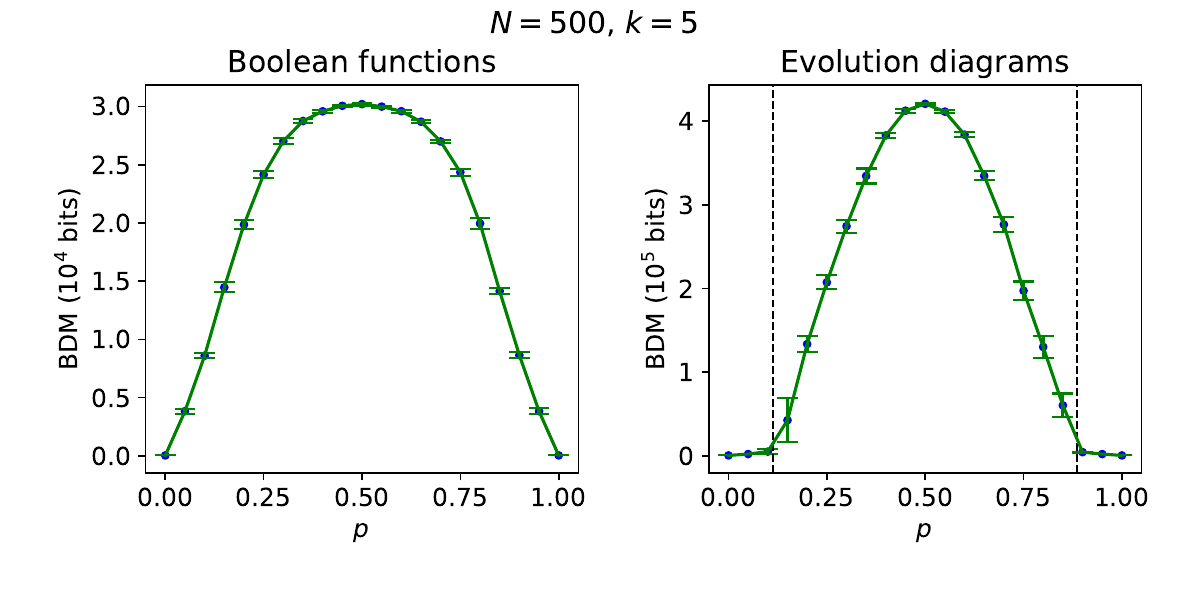}
    \caption{Top: randomness estimates of RBNs with $N=500$ and $k=5$ as a function of the parameter $p$. Bottom: estimates of average algorithmic complexity from samples of size 10. Vertical dotted lines represent the theoretical critical values of $p = 0.113$ and $p = 0.887$ at the edge of chaos. All RBNs share the same initial state and the same connectivity matrix, both generated randomly from uniform probability distributions.}
    \label{random_vs_p}
\end{figure}

Figure \ref{random_vs_p} presents results on simulations of RBNs with $N=500$ and $k=5$ that share the same initial state and the same connectivity matrix, both generated from uniform probability distributions. At the top of Figure \ref{random_vs_p} there are randomness estimates, for both the Boolean functions and the time evolution diagrams, as a function of parameter $p$. The results concerning the Boolean functions show ``causality gaps'' between the entropy/LZW estimates and the AC estimates using the BDM, and could be anticipated since the Boolean functions were constructed from a binomial probability distribution with parameter $p$ (see (\cite{e20080605})). The causality gaps are more noticeable in the randomness estimates of the time evolution diagrams; it can even be seen that the AC estimation is capable of revealing the transition to the chaotic regime on both sides of the ordered regime ($p = 0.113$ and $p = 0.887$), while both entropy and LZW are insensitive to such transition (beyond some small fluctuations) and give results from which it can only be suggested that the evolution diagrams arise from a binomial probability distribution. For greater statistical rigor, estimates of average algorithmic complexity from samples of size 10 of RBNs generated with a fixed value of p are shown at the bottom of Figure \ref{random_vs_p}; as in the upper part, we can see the transitions to the chaotic regime.

\begin{figure}
    \centering
    \includegraphics[width=0.64\linewidth]{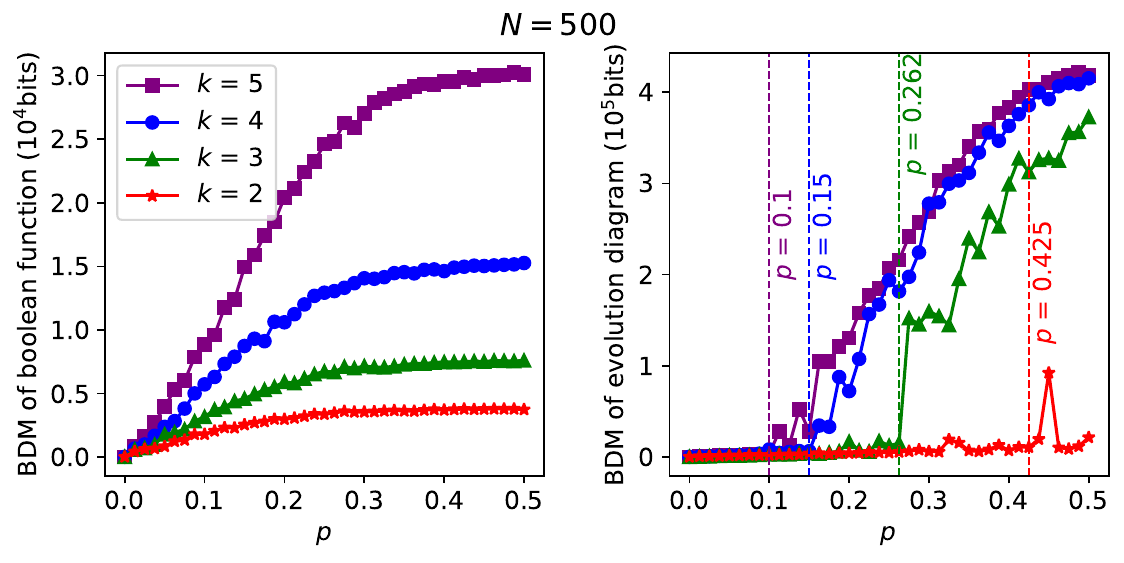}
    \caption{Left: Plots of boolean functions' BDM vs $p$. Right: Plots of time evolution diagrams' BDM vs $p$. Vertical dotted lines mark the critical $p$ values in each series. All RBNs share the same initial state, and all RBNs for a given in-degree $k$ share the same connectivity matrix; both types of objects were generated randomly from uniform probability distributions.}
    \label{BDMvsBDM}
\end{figure}

Next, Figure \ref{BDMvsBDM} shows some series of BDM (Boolean function) vs p, and BDM (evolution diagram) vs $p$, obtained for each specific value of in-degree $k$. In this case, $p$ takes on 41 values evenly spaced in the interval $\left[ 0.0,0.5 \right]$. Simulations were carried out with connectivity matrices generated randomly from uniform probability distributions, with the RBNs for a specific $k$ value sharing the same connectivity matrix. In each of the series, we can notice a sudden growth of the time evolution diagram's BDM starting from a certain $p$ value (marked with vertical lines). Such $p$ values are the critical $p$ values that point out the transition to the critical regime in each series, as can be confirmed by comparison with the theoretical $p$ values obtained from the chaos condition in RBNs (see Table \ref{tab1}). We can also notice that the critical $p$ value increases when going from one series to another with lower $k$, following the chaos condition in RBNs. Furthermore, as expected, in each series it is observed that an increase in BDM of the evolution diagrams goes hand in hand with an increase in BDM of the Boolean functions; it can also be observed that the series with greater $k$ have greater BDM values in the time evolution diagrams.

\begin{table}
    \caption{Comparison of critical $p$ values for the RBNs simulated in Figure \ref{BDMvsBDM}.}
	\centering
	\begin{tabular}{lll}
		\toprule
		$k$     & Critical $p$ value in Figure \ref{BDMvsBDM} & Theoretical Critical $p$ value \\
		\midrule
		5		& 0.100			& 0.113\\
4		& 0.150			& 0.146\\
3		& 0.262			& 0.211\\
2		& 0.425			& 0.50\\
		\bottomrule
	\end{tabular}
	\label{tab1}
\end{table}

\begin{figure}
    \centering
    \includegraphics[width=0.32\linewidth]{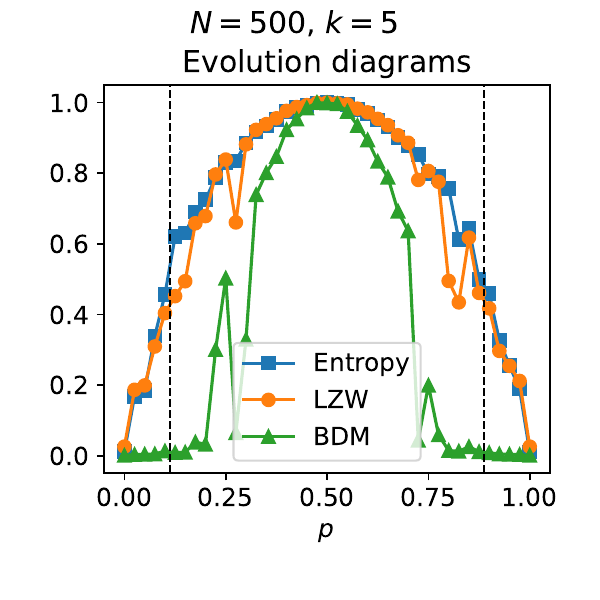}
    \caption{Randomness estimates of RBNs with $N=500$ and $k=5$ as a function of the parameter $p$. Vertical dotted lines represent the theoretical critical values of $p = 0.113$ and $p = 0.887$ at the edge of chaos. All RBNs share the same initial state and the same connectivity matrix. The connectivity matrix was generated randomly from a binomial probability distribution with $p=0.5$ and $n=499$.}
    \label{random_vs_p_B}
\end{figure}

The results presented up to this point were obtained from simulations that consider connectivity matrices generated from uniform probability distributions. However, it is possible that non-uniform probability distributions introduce a change in the dynamics and therefore also introduce changes in the critical points. Although considerations about non-uniform distributions are beyond the objectives of this work, Figure \ref{random_vs_p_B} shows the results obtained by repeating the simulations carried out for Figure 3, but this time the connectivity matrix was generated from a binomial probability distribution with $P=0.5$ and $n = 499$ that is, the nodes of the Boolean network are labeled from $0$ to $499$ and the quantity that follows the binomial distribution is the probability that one of these nodes is chosen to determine each of the entries in the connectivity matrix; the condition $P=0.5$ implies that the nodes with the highest probability of being chosen are nodes 249 and 250). In this case, transitions towards the chaotic regime continue to be observed and these can be revealed more noticeably by AC estimates; although, unlike the results presented in Figure 3, the critical values of $p$ do not coincide with the theoretical ones (indicated with dotted lines in Figure \ref{random_vs_p_B}) expected from a connectivity matrix constructed from a uniform probability distribution. The latter shows a change in the dynamics and the critical points when taking into account non-uniform probability distributions.

\subsection{RBNs with $N=10$ and $k=5$}
\label{pert_analysis}

\begin{figure}
    \centering
    \includegraphics[width=\linewidth]{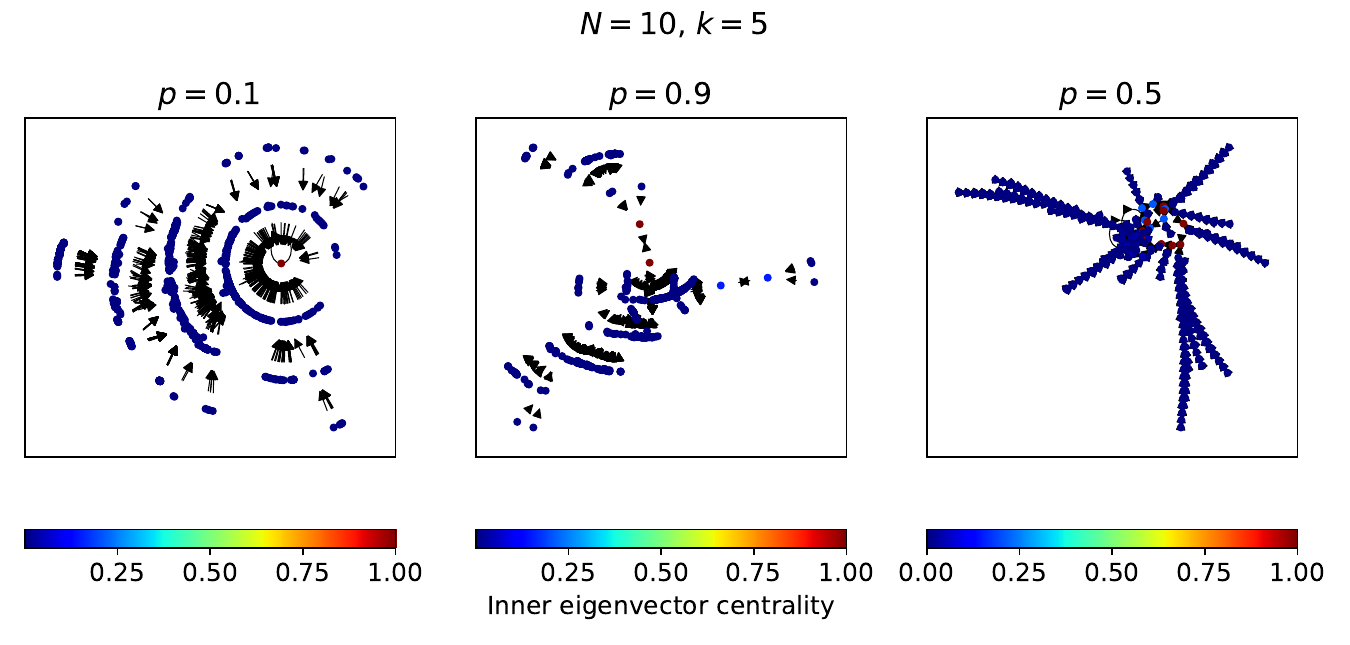}
    \caption{Transition diagrams for three RBNs with 10 nodes and \textit{in-degree} 5. The transition diagrams correspond to the values $p=0.1$, $p=0.9$ and $p=0.5$. The color of the nodes denotes their inner eigenvector centrality.}
    \label{transitions}
\end{figure}

In this subsection, some RBNs with 10 nodes and \textit{in-degree} of 5 are analyzed from their complete transition diagram - which have $2^{N} = 1024$ possible states. In Figure \ref{transitions} there are three transition diagrams, two for RBNs corresponding to the opposite values $p=0.1$ and $p=0.9$, and one for an RBN with a value of $p = 0.5$. The \textit{internal eigenvector centrality} measure or \textit{prestige} of the nodes (\cite{e2a71e41-68e3-385f-b794-7c4c0555a2bb,newman2018networks}) in each diagram was calculated. In the transition diagram of the RBN with $p = 0.1$ a single fixed point attractor is observed, in which all states fall after a few time steps. In the transition diagram of the RBN with $ p = 0.9$ a similar behavior is observed with two limit cycle attractors. On the other hand, in the RBN with $p = 0.5$ several attractors with more time steps are observed. It can be noted that the attractors have the greatest prestige in each transition diagram.

The transition diagrams obtained are conducive to performing perturbation analysis. In this work, perturbations were carried out on our RBNs according to the following procedure:

\begin{enumerate}
     \item Perform randomness measurements of the unperturbed transition diagram.
     \item Create a list of nodes with the highest prestige.
     \item Perturbation: locate the node with the highest prestige in the list, disconnect it from the transition diagram, and perform randomness measurements.
     \item Remove the most prestigious node from the list.
     \item Restore disconnected node to transition diagram.
     \item Repeat steps 3, 4, and 5 until the list of most prestigious nodes is exhausted.
\end{enumerate}

Let's call this perturbation method \textit{perturbation of most relevant nodes}. Similarly, let's call \textit{perturbation of less relevant nodes} the perturbation method where nodes with lower prestige are disconnected (modifying step 2).

\begin{figure}
    \centering
    \includegraphics[width=\linewidth]{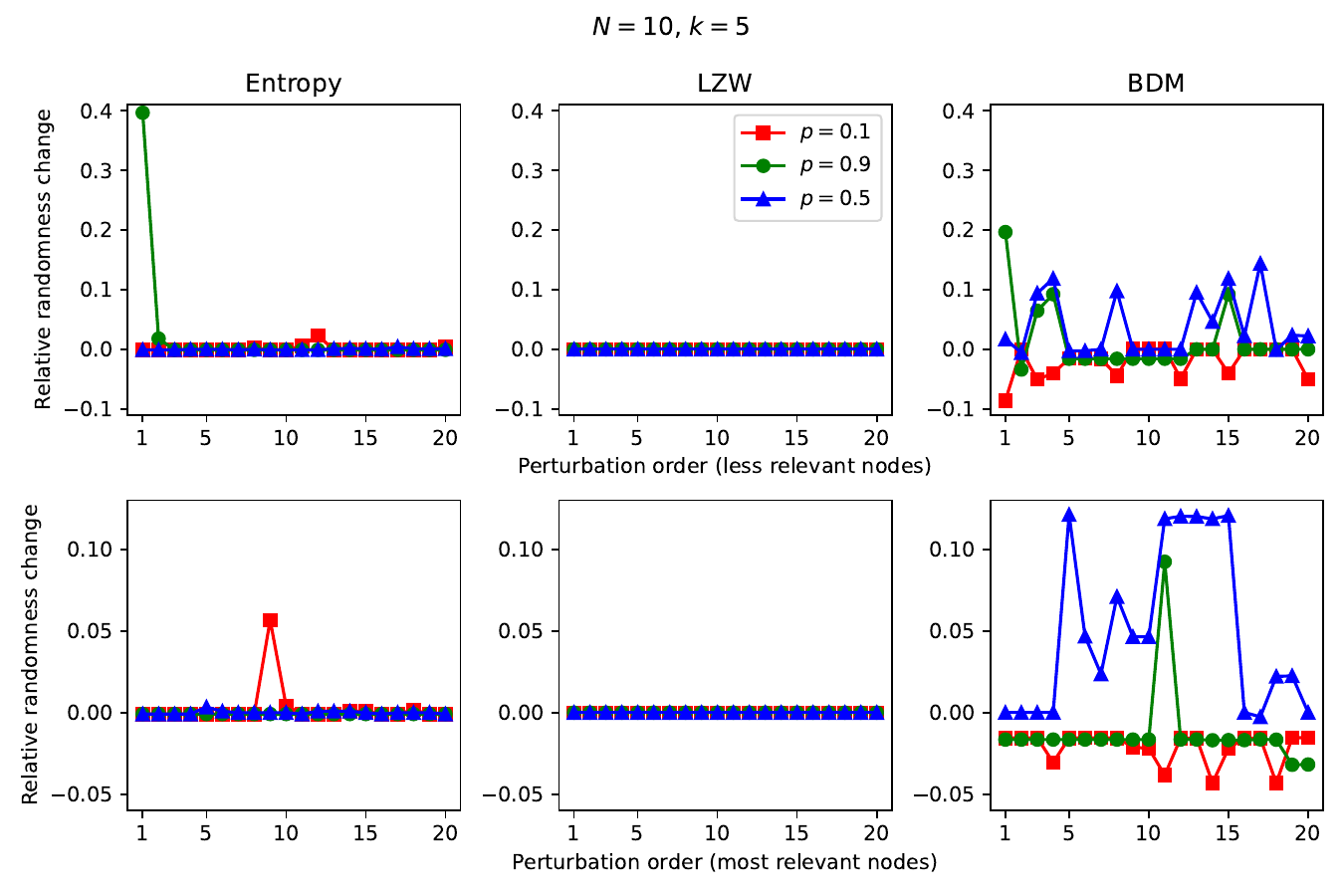}
    \caption{Relative randomness change produced by perturbations of the transition diagrams from Figure \ref{transitions}. The randomness measures were applied to the adjacency matrices of the transition diagrams.}
    \label{perturbations}
\end{figure}

%\begin{figure}[t]
%    \centering
    %\includegraphics[width=0.64\linewidth]{figures/AID.pdf}
    %\caption{Algorithmic information dynamics resulting from perturbation of less (left) and most (right) relevant nodes in the transition diagrams from Figure \ref{transitions}.}
    %\label{AID}
%\end{figure}

Figure \ref{perturbations} shows the results of performing the aforementioned perturbations. The 20 most relevant-prestigious nodes and the 20 least relevant-prestigious nodes in each transition diagram in Figure \ref{transitions} were perturbed. In each perturbation, the \textit{relative randomness change} (defined here as $\left[ R\left( G \right) - R\left( G \backslash e \right) \right]/R\left( G \right)$ ) of the respective transition diagram was calculated from its adjacency matrix. The perturbations series obtained show that the compressibility by LZW is blind to the perturbations, as is the entropy (except for a few perturbations), while the algorithmic complexity - estimated by BDM - is sensitive in the face of these perturbations. 

In the graphs at the right of Figure \ref{perturbations}, values of relative shared mutual algorithmic information (AID) between the perturbed and unperturbed transition diagrams of Figure \ref{transitions} are shown. Here, we can observe that for the transition diagram with $p = 0.1$ the first least relevant node makes a relatively high causal contribution to the transition diagram (because $\textrm{AID} < 0$ for that perturbation). This is to be expected since the least relevant nodes are found on the periphery of the transition diagram to which they belong, which makes them more preceding nodes in the temporal evolution of the respective RBN. Regarding the perturbation of more relevant nodes, we see a few AID inverted peaks under a constant low absolute value; this tells us that the majority of less relevant nodes are not essential in the explanation of the system, so it is likely that they are elements produced by the natural course of evolution of the system. The latter is to be expected since the most relevant-prestigious nodes are usually close to the attractors. A similar analysis follows from the results for the transition diagram with $p = 0.9$, although in this case the first least relevant node does not take part in the algorithmic content of the transition diagram (because $\textrm{AID} > 0$). Finally, for the transition diagram with $p = 0.5$, there are more nodes with low causal contribution ($\textrm{AID} > 0$) with respect to the other two transition diagrams; the existence of such nodes with low causal contribution is perhaps due to the fact that the relevant nodes were selected from among the several basins of attraction in the transition diagram, which are separated from each other.

\section{Conclusions}
We used three measures of randomness to analyze the dynamics of RBNs with synchronous updates. Two main results were shown in this work: a) the randomness estimation by BDM is capable of showing jumps from null values of randomness and ``causality gaps'', which makes BDM the randomness measure capable of detecting regime changes in a more noticeable way with respect to the other measures of randomness; b) BDM is the only randomness measure that is capable of revealing the causal contribution of certain states according to the order they occupy in the temporal evolution of an RBN. The theoretical results presented in subsection \ref{pert_analysis} introduce the application of reprogrammability analysis in Boolean networks, potentially leading to practical applications. Specifically, we believe that a future work may be the extension of the analysis to RBN models with asynchronous updates (\cite{CHAVES2005431}), or the application of perturbation analysis on connectivity graphs in regulatory network models to help simplify such graphs and their Boolean rules/functions ---i.e., to perform gene knock-outs in order to find the causal regulatory network.

Finally, a brief explanation is in order for the accuracy of the Block Decomposition Method. For larger networks, some effects may be lost, but precisely the sensitivity of the method holds for small perturbations which popular lossless compression would fail (\cite{e20080605}).

\bibliographystyle{unsrtnat}
\bibliography{references}  %%% Uncomment this line and comment out the ``thebibliography'' section below to use the external .bib file (using bibtex) .





% \end{thebibliography}

\end{document}